\documentclass[11pt]{article}
\usepackage[T1]{fontenc}
\usepackage{graphics}

\def\sinc{\mathop{\rm sinc}\nolimits}
\def\sech{\mathop{\rm sech}\nolimits}
\begin{document}

\title{Gravitational Radiation and the Small-Scale Structure of Cosmic Strings}

\author{Xavier Siemens and Ken D. Olum}

\maketitle
{\centering \par \emph{Institute of Cosmology}\par{}\par}

{\centering \par\emph{Department of Physics and Astronomy}\par{}\par}

{\centering \par\emph{Tufts University, Medford MA 02155, USA}\par{}\par}

\begin{abstract}
We calculate the gravitational radiation emitted by an infinite cosmic
string with two oppositely moving wave-trains, in the small amplitude
approximation.  After comparing our result to the previously studied
cases we extend the results to a new regime where the wavelengths of
the opposing wave-trains are very different. We show that in this case
the amount of power radiated vanishes exponentially. This means that
small excitations moving in only one direction may be very long lived,
and so the size of the smallest scales in a string network might be much
smaller than what one would expect from gravitational back reaction.
This result allows for a potential host of interesting cosmological
possibilities involving ultra-high energy cosmic rays, gamma ray
bursts and gravitational wave bursts.
\end{abstract}

\section{Introduction }

Topological defects are a prediction of most particle physics models
that involve symmetry breaking and are therefore quite generic. They are
formed when the topology of the vacuum manifold of the low energy
theory is non-trivial \cite{1}. The typical size of regions which
acquire different vacuum expectation values depends on the type of
phase transition but it can never exceed the horizon size, due to
causality. The specific defect formed in turn depends on the symmetry
of the groups involved, in particular on the homotopy group of the
vacuum manifold.  If the vacuum manifold contains disconnected parts
domain walls are formed, if it contains unshrinkable loops strings are
formed and if it contains unshrinkable spheres monopoles are
formed. More complicated hybrid defects may be formed if there is more
than one phase transition. For a review see \cite{2}.

The most simple phase transition where cosmic strings are produced is
\begin{equation}
\label{u1break}
U(1)\stackrel{\eta _{s}}{\longrightarrow }1\, .
\end{equation}
The vacuum manifold is \( U(1) \), which contains unshrinkable
loops.
The phase transition, occurring at energy \( \eta _{s} \), traps the magnetic
flux of the gauge fields associated with the \( U(1) \) symmetry into strings
of mass per unit length \( \mu \sim \eta ^{2}_{s} \). The resulting system
is a network of long strings and closed loops. Numerical simulations and analytic
studies show that this network quickly evolves into a ``scaling regime''
(see \cite{2} and references therein), where the energy density of the string
network is a constant fraction of the radiation or matter density and the
statistical properties of the system such as the correlation lengths of long
strings and average sizes of loops scale with the cosmic time \( t \). The
question of the size of the small-scale structure of cosmic string networks,
however, has so far resisted concordance. Simulations show that most loops and
the wiggles on the long strings have the smallest possible size, the simulation
resolution. However, the prevailing opinion on this issue is that the size of
small-scale structure also scales with the cosmic time \( t \) and its value
is given by the gravitational back-reaction scale \( \Gamma G\mu t \), where
\( \Gamma  \) is a number of order 100 and \( G \) is Newton's constant.
This possibility was first pointed out in \cite{3}.

Cosmic strings have been a popular choice in the literature because unlike monopoles
and domain walls they do not cause cosmological disasters. Indeed, they originally
were considered good candidates for structure formation. Recent cosmological
data, in particular the cosmic microwave background power spectrum, is not consistent
with density fluctuations produced only by strings.  A combination
of density fluctuations produced by inflation and strings is possible,
and strings with small values of $\mu$ would have little gravitational
effect. Cosmic strings are also good candidates
for a variety of interesting cosmological phenomena such as gamma ray
bursts \cite{4}, gravitational wave bursts \cite{5} and ultra high energy
cosmic rays \cite{6}. However, some of the predictions of these models depend
sensitively on the so far unresolved question of the size of the small-scale 
structure.

Sakellariadou \cite{Maria} computed the radiation from a cosmic string
with helical standing waves, which one can consider to be composed of
excitations of equal wavelength traveling in the two directions.
Hindmarsh \cite{7} calculated the radiation from colliding wiggles of
different wavelengths on a straight string, in a low-amplitude
approximation, and found that the radiation rate from small wavelength
wiggles approached a constant as the wavelength of the wiggles in the
opposite direction became large.  However, Garfinkle and Vachaspati
\cite{Vachaspati} showed that a wiggle moving on a straight string
does not radiate.  This poses a puzzle because an excitation of
sufficiently large wavelength would seem to appear straight to a tiny
wiggle passing over it.

With this work we hope to shed some light on these issues. The following
section contains a review of string motion and a derivation of the
energy emitted in the form of gravitational radiation per unit solid
angle. Section 3 starts with a derivation of the power emitted from
infinite planar excitations on a string with a subsequent
specialisation to sinusoidal wave-trains in the case where the
amplitude is small. We show that in the similar wavelength case the
result of Hindmarsh \cite{7} applies and how an argument on the limit
on the size of the small-scale structure consistent with \cite{3} can be
obtained from this result.  We then demonstrate that this argument
cannot generally be made and in particular show that when the
wavelengths are very different the radiation vanishes
exponentially. In section 4, we discuss these results and their relevance to the
question of the small-scale structure in cosmic string networks and
conclude with some remarks on the effect of these results on
cosmological phenomena.

\section{Preliminaries}

When the typical length scale of a cosmic string is much larger than its thickness,
\( \delta _{s}\sim \eta ^{-1}_{s} \), and long-range interactions between different
string segments can be neglected, the string can be accurately modeled by a one
dimensional object. Such an object sweeps out a two dimensional
surface in space-time referred
to as the string world-sheet. 

The infinitesimal line element in Minkowski space-time with metric \( \eta _{\mu \nu }={\textrm{diag}}(1,-1,-1,-1) \)
is
\begin{equation}
\label{lineelem}
ds^{2}=\eta _{\mu \nu }dx^{\mu }dx^{\nu }=\eta _{\mu \nu }
x^{\mu }_{,a}x_{,b}^{\nu }d\zeta ^{a}d\zeta ^{b},
\end{equation}
where \( a=0,1 \) labels the two internal parameters of the string world-sheet
and \( x_{,a}^{\mu }=\partial x^{\mu }/\partial \zeta ^{a} \). One can then write
the induced metric on the world-sheet of the string as 
\begin{equation}
\label{inducedmetric}
\gamma _{ab}=\eta _{\mu \nu }x^{\mu }_{,a}x_{,b}^{\nu }\, .
\end{equation}

For an infinitely thin string we can use the Nambu-Goto action. It is proportional
to the invariant area swept by the string,
\begin{equation}
\label{stringaction}
S=-\mu \int dA=-\mu \int d^{2}\zeta \sqrt{-\gamma },
\end{equation}
where \( \gamma =\det(\gamma _{ab}) \) and \( \mu  \) is the mass per unit
length of the string. The parametrisation-invariant energy-momentum tensor for
a cosmic string is given by
\begin{equation}
\label{TmunuGaugeInv}
T^{\mu \nu }(x)=\mu \int d^{2}\zeta \sqrt{-\gamma }
\gamma ^{ab}x_{,a}^{\mu }x_{,b}^{\nu }\delta ^{(4)}(x-X(\zeta )).
\end{equation}
In the light-cone gauge the metric takes the form
\begin{equation}
\label{gammaab}
\gamma _{ab}=\sqrt{-\gamma }\left( \begin{array}{cc}
0 & 1\\
1 & 0
\end{array}\right) ,\: \gamma ^{ab}=\frac{1}{\sqrt{-\gamma }}\left( \begin{array}{cc}
0 & 1\\
1 & 0
\end{array}\right) 
\end{equation}
because the arc element is \( ds^{2}=2x_{,u}
\cdot x_{,v}dudv \) with \( x^{\mu }=x_{R}^{\mu }(u)+x_{L}^{\mu }(v) \),
and \( {x'}_{R}^{\mu }(u) \) and \( {x'}_{L}^{\mu }(v) \) are both null vectors.
In this gauge the stress energy tensor (\ref{TmunuGaugeInv}) can be put in
the form
\begin{equation}
\label{TmunuLightConeGauge}
T^{\mu \nu }(x)=\mu \int dudv(x_{,u}^{\mu }x_{,v}^{\nu }
+x_{,u}^{\nu }x_{,v}^{\mu })\delta ^{(4)}(x-x(\zeta )).
\end{equation}

Far from a source localized in space and time, the total energy
radiated in the form of gravity waves in the direction of \( {\bf k}
\) is \cite{8}
\begin{equation}
\label{weinberggravrad}
\frac{dE}{d\Omega }=2G\int ^{\infty }_{0}d\omega \omega ^{2}
\{T^{\mu \nu *}(k)T_{\mu \nu }(k)-\frac{1}{2}
\left| T^{\mu }_{\mu }(k)\right| ^{2}\}.
\end{equation}

So what we need is the Fourier transform of the stress energy tensor
\begin{equation}
\label{TmunuFT}
T^{\mu \nu }(k)=\frac{1}{2\pi }\int d^{4}xT^{\mu \nu }(x)e^{ik\cdot x}
\end{equation}
which is given by
\begin{equation}
\label{TmunuLightConeGaugeFT}
T^{\mu \nu }(k)=\frac{1}{2\pi }\mu \int 
d^{4}xe^{ik\cdot x}\int dudv(x_{,u}^{\mu }x_{,v}^{\nu }+x_{,u}^{\nu }x_{,v}^{\mu })
\delta ^{(4)}(x-x(\zeta )).
\end{equation}

Using \( x^{\mu }=[a^{\mu }(u)+b^{\mu }(v)]/2 \) yields for (\ref{TmunuLightConeGaugeFT}),
\begin{equation}
\label{Tmunuforstring}
T^{\mu \nu }(k)=\frac{1}{8\pi }\mu 
\int dudv(a'^{\mu }b'^{\nu }+a'^{\nu }b'^{\mu })e^{ik\cdot (a+b)/2}.
\end{equation}
Following \cite{7} this expression can in turn can be written as
\begin{equation}
\label{TmunuSimplified}
T^{\mu \nu }(k)=\frac{1}{8\pi }\mu (A^{\mu }(k)B^{\nu }(k)+A^{\nu }(k)B^{\mu }(k))
\end{equation}
where
\begin{equation}
\label{AandBofk}
A^{\mu }(k)=\int ^{+\infty }_{-\infty }d\xi 
a'^{\mu }(\xi )e^{ik\cdot a(\xi )/2},\: B^{\mu }(k)=\int ^{+\infty }_{-\infty }
d\xi b'^{\mu }(\xi )e^{ik\cdot b(\xi )/2}.
\end{equation}
We can then re-write the total energy in the direction 
of \( {\bf k} \) (\ref{weinberggravrad}) as
\begin{equation}
\label{weinberggravradforstrings}
\frac{dE}{d\Omega }=\frac{1}{16\pi ^{2}}G\mu ^{2}\int ^{\infty }_{0}
d\omega \omega ^{2}\{\left| A\right| ^{2}\left| B\right| ^{2}
+\left| A^{*}\cdot B\right| ^{2}-\left| A\cdot B\right| ^{2}\}.
\end{equation}

\section{Gravitational Radiation from Cosmic Strings}

\subsection{Planar Excitations}

\begin{figure}
{\centering \resizebox*{0.8\columnwidth}{0.1\textheight}{\includegraphics{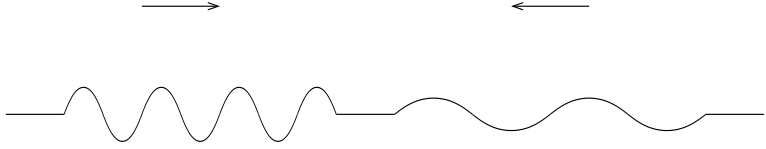}} \par}
\caption{Two oppositely moving wave-trains on an infinite straight cosmic string. We
ignore the effects of the kinks at the edges of the wave-trains because later
we will be taking the length of the wave-trains to \protect\( \infty
\protect \).}
\label{fig:wiggles}
\end{figure}

We consider planar string excitations moving on the $z$-axis in 
the two directions as shown in Figure \ref{fig:wiggles}. 

We are free to take
\begin{equation}
\label{aplanarexcit}
a'^{\mu }(u)=(1,f'(u),0,-\sqrt{1-f'^{2}}),\: 
a^{\mu }(u)=\left( u,f(u),0,-\int \sqrt{1-f'^{2}}du\right) 
\end{equation}
where the sign of the square root is chosen such that $u$ decreases in the positive $z$ direction,
which yields from (\ref{AandBofk})
\begin{equation}
\label{amuofk}
A^{\mu }(k)=\int _{-\infty }^{\infty }du\, 
a'^{\mu }(u)\exp \frac{i}{2}\left[ \omega u-k_{x}f
+k_{z}\int \sqrt{1-f'^{2}}du\right] 
\end{equation}
exactly. The situation for \( B^{\mu }(k) \) is completely
analogous.

Equation (\ref{weinberggravradforstrings}) gives the energy radiated
from a localized interaction.  In Appendix \ref{appendix:periodic}, we
extend this to the power per unit length from an infinite periodic
wave train.
We find that the radiation is emitted in a discrete set of cones and
at a discrete set of frequencies.  Specifically, let $\Delta_a$ be
the average of $|a'_z|$ ,
\begin{equation}
\label{Deltaa}
\Delta _{a}=\frac{1}{\lambda _{a}}\int _{0}^{\lambda _{a}}du\, \sqrt{1-f'^{2}}
\end{equation}
where $\lambda_a$ is the wavelength of the wiggles of $a^{\mu }(u)$, and let
\begin{equation}
\kappa_a ={2\pi\over\lambda_a}
\end{equation}
and similarly for $\Delta_b$ and $\kappa_b$.  Then
for each set of
positive integers $n, m$ satisfying
\begin{equation}
\label{generalsumlimits}
(1-\Delta _{a})/(1+\Delta _{b})<|n\kappa _{a}/m\kappa _{b}|<(1+\Delta _{a})/(1-\Delta _{b}).
\end{equation}
we have radiation with polar angle $\theta$ given by
\begin{equation}
\label{theta}
\cos \theta =
\frac{n\kappa _{a}-m\kappa _{b}}{n\kappa _{a}\Delta _{b}+m\kappa _{b}\Delta _{a}}\,,
\end{equation}
at frequency
\begin{equation}
\label{omega}
\omega =2\frac{n\kappa _{a}\Delta _{b}+m\kappa _{b}\Delta _{a}}{\Delta _{a}
+\Delta _{b}}\,,
\end{equation}
(so that $\omega +\Delta_a k_z = 2n\kappa_a$ and $\omega -\Delta_bk_z = 2m\kappa_b$)
with power per unit length per unit azimuthal angle
\begin{equation}
\label{dPdz}
\frac{dP}{dzd\phi }=\frac{G\mu ^{2}}{\Delta _{a}+\Delta _{b}}
\sum_{n,m} (n\kappa _{a}+m\kappa _{b})|A_{n}|^{2}|B_{m}|^{2}
\end{equation}
where
\begin{eqnarray}
\label{Anudef}
A_n^\mu &=&{1\over\lambda_a}\int _{-\lambda _{a}/2}^{\lambda _{a}/2}du\,
a'^{\mu }e^{ik\cdot a(u)/2}\nonumber\\
&=&{1\over\lambda_a}\int _{-\lambda _{a}/2}^{\lambda _{a}/2}du\,
a'^{\mu }\exp{i\over 2}\left[\omega u+k_z\int
_{0}^{u}\sqrt{1-f'^{2}}du'-k_{x}f(u) \right]
\end{eqnarray}

If we fix the shape of $a^\mu$, but vary the scale so that $\lambda_a$
becomes large, then $\Delta_a$ is fixed and $\kappa_a\to0$.  If we
follow a single mode $n, m$ as we go
toward this limit, \( \theta \) reaches \( \pi \) at a finite value
of $\kappa_a$, given by
\begin{equation}
\label{kappaamax}
\kappa ^{m}_{a}=\frac{m}{n}\kappa _{b}(1-\Delta _{a})/(1+\Delta _{b}).
\end{equation}
For larger values of $\kappa_a$, the given mode does not exist.

\subsection{Sinusoidal waves }

\subsubsection{Small amplitudes}

We will consider excitations whose amplitude is small compared to their
wavelength, so that $f'\ll1$.  This enables us to replace
$\sqrt{1-f'^2}$ by its average, $\Delta_a$, in (\ref{Anudef}), to get
\begin{equation}
\label{Amusubnapprox}
A^{\mu }_{n}\approx \frac{1}{\lambda _{a}}\int _{-\lambda _{a}/2}^{\lambda _{a}/2}du\, 
a'^{\mu }(u)\exp i\left[ n\kappa _{a}u-k_{x}f(u)/2\right] .
\end{equation}

In particular if we take the wave trains to be sinusoidal, namely,
\begin{equation}
\label{sinusoidalwiggles}
f'(u)=\epsilon _{a}\cos (\kappa _{a}u),\: 
f(u)=\frac{\epsilon _{a}}{\kappa _{a}}\sin (\kappa _{a}u)
\end{equation}
with \( \epsilon _{a}\ll 1 \), and thus
\begin{equation}\label{Deltaapprox}
 \Delta _{a}\approx 1-\epsilon ^{2}_{a}/4\,,
\end{equation}
for the time component we have
\begin{equation}
\label{Azero}
A^{0}_{n}\approx \frac{1}{\lambda _{a}}\int _{-\lambda _{a}/2}^{\lambda _{a}/2}du
\, \exp i\left[ n\kappa _{a}u-\epsilon _{a}k_{x}\sin (\kappa _{a}u)/2\kappa _{a}\right] 
\end{equation}
which can be integrated to get
\begin{equation}
\label{AzeroBess}
A^{0}_{n}\approx J_{n}\left( \frac{\epsilon _{a}k_{x}}{2\kappa _{a}}\right) 
\end{equation}
where \( J_{n} \) is the Bessel function  of order \( n \).
The first component can be put in terms of \( A_{0} \) using integration by
parts
\begin{equation}
\label{Aone}
A^{1}_{n}(k)=-\frac{2n\kappa _{a}}{k_{x}}A^{0}_{n}
\end{equation}
and the third component can be also be written in terms of \( A_{0} \) as
\begin{equation}
\label{Athree}
A^{3}_{n}=\Delta _{a}A^{0}_{n}
\end{equation}
so that
\begin{equation}
\label{Ansquared}
\left| A_{n}\right| ^{2}\approx \left( 1-\frac{4n^{2}\kappa ^{2}_{a}}{k^{2}_{x}}
-\Delta _{a}^{2}\right) J^{2}_{n}\left( \frac{\epsilon _{a}k_{x}}{2\kappa _{a}}\right) .
\end{equation}

A similar analysis for \( B \) and using (\ref{Deltaapprox}) yields the 
expression for the power in terms of Bessel functions,
{\arraycolsep=0pt
\begin{eqnarray}
\label{fullpow}
\frac{dP}{dzd\phi }\approx \frac{G\mu ^{2}}{\Delta _{a}+\Delta _{b}}
\sum _{n,m}&&(n\kappa _{a}+m\kappa _{b})
\left( \frac{\epsilon ^{2}_{a}}{2}
-\frac{4n^{2}\kappa ^{2}_{a}}{k^{2}_{x}}\right) \left( \frac{\epsilon ^{2}_{b}}{2}
-\frac{4m^{2}\kappa ^{2}_{b}}{k^{2}_{x}}\right)
\nonumber\\
&&\mbox{}\times J^{2}_{n}\left( \frac{\epsilon _{a}k_{x}}
{2\kappa _{a}}\right) J^{2}_{m}\left( \frac{\epsilon _{b}k_{x}}{2\kappa _{b}}\right) 
\end{eqnarray}}

\begin{figure}
{\centering \resizebox*{0.35\columnwidth}{0.35\textheight}{\includegraphics{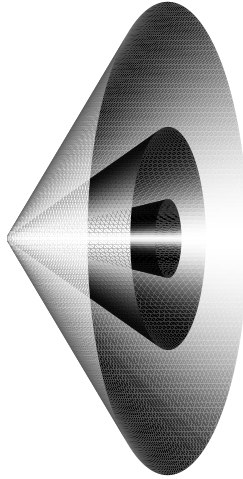}} \par}
\caption{Discrete radiation cones given by (\ref{theta}) and (\ref{fullpow}) 
for $m=1$ and $n=2,4$ and $20$ with 
$\epsilon_a ^2=\epsilon_b ^2 =0.1$ and $\kappa_a = \kappa_b$. Shading denotes the intensity of the radiation, with white being most
intense.  As $n$ gets larger, the cone becomes narrow and the
radiation gets concentrated in a narrow band around where the cone
intersects the plane of the string.}
\label{fig:cones}
\end{figure}

As an example, Figure \ref{fig:cones} shows some of the discrete radiation cones given by 
(\ref{theta}) 
with power calculated from (\ref{fullpow}).
 
\subsubsection{Similar wavelength wiggles and the back-reaction argument}

If \( \kappa _{a} \) and \( \kappa _{b} \) are similar in magnitude, 
and \( \epsilon _{a},\epsilon _{b}\ll 1 \), then the arguments of both Bessel 
functions will be much less than 1. In this case we can use the 
small-argument approximation for the Bessel functions \cite{9}, 
\begin{equation}
\label{BessSmallArgExp}
J_{n}(x)\approx \frac{x^{n}}{2^{n}n!}
\end{equation}
and so we see that the dominant contribution to (\ref{fullpow})
comes from the \( n=m=1 \) mode, and the power can be written
\begin{equation}
\label{dPdzdphismallargapprox}
\frac{dP}{dzd\phi }\approx \frac{G\mu ^{2}\epsilon ^{2}_{a}\epsilon ^{2}_{b}}{32}
(\kappa _{a}+\kappa _{b})\,.
\end{equation}

Integrating over \( \phi  \) yields
\begin{equation}
\label{hindmarshpow}
\frac{dP}{dz}\approx \frac{1}{16}\pi G\mu ^{2}\epsilon ^{2}_{a}\epsilon ^{2}_{b}
(\kappa _{a}+\kappa _{b})
\end{equation}
This agrees with the result of \cite{7}.  The factor of 16 results
from our choice of normalization in the definitions of \( A^{\mu }(k) \)
and \( B^{\mu }(k) \).

An argument that can be thought to follow from this result is the
following.
Consider short wavelength excitations in $b$ interacting with long
wavelength excitations in $a$.  If we put
\begin{equation}
\label{badapprox}
\kappa _{a}\ll \kappa _{b}
\end{equation}
in (\ref{hindmarshpow}) we obtain
\begin{equation}
\label{approxhindpow}
\frac{dP}{dz}\sim \frac{1}{8}\pi ^{2}G\mu ^{2}\epsilon_a^2\epsilon ^{2}_{b}\frac{1}{\lambda _{b}}
=\frac{1}{8}\pi ^{2}G\mu \epsilon_a^2\frac{\delta \mu }{\lambda _{b}}
\end{equation}
where \( \delta \mu \sim \epsilon ^{2}_{b}\mu  \) is the contribution of the
short wavelength wiggles to the mass per unit length of the string.
We expect the radiated
energy to be taken from the small-scale wiggles, so we expect 
\begin{equation}
\label{radpowfromsmall}
\frac{d(\delta \mu )}{dt}\sim -\frac{dP}{dz}\sim -\frac{1}{8}\pi ^{2}G\mu 
\epsilon_a^2\frac{\delta \mu }{\lambda _{b}}
\end{equation}
which has the solution 
\begin{equation}
\label{extramassdep}
\delta \mu \propto e^{-t/\tau }
\end{equation}
with 
\begin{equation}
\label{typicaltime}
\tau =\frac{8\lambda _{b}}{\pi ^{2}\epsilon_a^2G\mu }.
\end{equation}

If we assume that large wavelength wiggles exist with $\epsilon\sim1$, we would
conclude by the above argument that the minimum wavelength of wiggle
that can survive until the present day is
\begin{equation}
\label{ssslimit}
\lambda_{\rm min}\sim G\mu t_0
\end{equation}
where $t_0$ is the present age of the universe.  On smaller scales,
wiggles are exponentially suppressed.

If we extend this analysis to a loop, the structure
of the loop that enables it to be closed must involve features
whose amplitude is comparable to their wavelength, so the condition
$\epsilon\sim1$ is always met.

However, it turns out that (\ref{hindmarshpow}) is never correct in
the regime of (\ref{badapprox}) with $\epsilon_a\sim1$.  Even \(
\epsilon _{a},\epsilon _{b}\ll 1 \) is not sufficient, because we
might have, for example, \( k_{x}/\kappa _{a}\gg 1 \), and then the
argument of $J_n$ in (\ref{fullpow}) would not be small.  Explicitly,
\begin{eqnarray}
\label{kx}
k_{x}&=&\omega \sin (\theta )\cos (\phi )
\nonumber
\\
&=
&\frac{\cos \phi }{\Delta _{a}+\Delta _{b}}\sqrt{(\Delta _{b}n\kappa _{a}
+\Delta _{a}m\kappa _{b})^{2}-(n\kappa _{a}-m\kappa _{b})^{2}}
\end{eqnarray}
so \( k_{x} \) is at most of order \( \sqrt{nm\kappa
_{a}\kappa _{b}} \) and thus we are in the small-argument regime whenever
\begin{equation}
\label{goodapprox}
\epsilon _{a}^{2}\kappa _{b}/\kappa _{a}\ll 1,
\: \epsilon _{b}^{2}\kappa _{a}/\kappa _{b}\ll 1.
\end{equation}
Larger values of \( n \) and \( m \) could in principle make the arguments
of the Bessel functions larger, but it turns out that in this situation the
contribution is dominated by that of the lowest mode.

One can write (\ref{goodapprox}) in terms of the wavelengths of the sinusoids,
\( \lambda _{a,b}=2\pi /\kappa _{a,b} \) and their amplitudes, 
\( A_{a,b}=\epsilon _{a,b}/\kappa _{a,b} \),
as
\begin{equation}
\label{AaAb}
A_{a},A_{b}\ll \sqrt{\lambda _{a}\lambda _{b}}
\end{equation}
which is invariant under Lorenz boosts along the string. Rather than
merely requiring that the amplitude of each wave train be small
compared to its own wavelength to have the simple case, we must
require that each amplitude be small as compared to the geometric mean
wavelength. Note that these conditions cannot hold if
$\lambda_a\gg\lambda_b$ and $\epsilon_a\sim1$. As an example of this,
in the case of small ``chiral'' wiggles (those moving in one direction
only) on a loop, the conditions (\ref{goodapprox}) don't hold and we
expect wiggles to survive on the loop longer than (\ref{typicaltime}).

It should be noted that if both of the wiggles have about the same wavelengths
then an argument analogous to the one above can be made \cite{10}. In this
case wiggles in both directions contribute to the radiation power, and
the extra energy in the string declines as $1/t$.

\subsubsection{Chiral excitations}

In the following we take the small argument limit for the \( B \) part of (\ref{fullpow}),
namely \( \epsilon _{b}^{2}\kappa _{a}/\kappa _{b}\ll 1 \), 
or $A_{b}\ll \sqrt{\lambda _{a}\lambda _{b}}$,
but do not use that
approximation for the \( A \) part of the expression. This corresponds to the
situation where the opposing excitations have very different wavelengths. Thus
(\ref{fullpow}) becomes
\begin{equation}
\label{powbess}
\frac{dP}{dzd\phi }=\frac{G\mu ^{2}\epsilon ^{2}_{b}}{8}\sum _{n}
(n\kappa _{a}+\kappa _{b})\left( \frac{4n^{2}\kappa ^{2}_{a}}{k^{2}_{x}}
-\frac{\epsilon ^{2}_{a}}{2}\right) J^{2}_{n}\left( 
\frac{\epsilon _{a}k_{x}}{2\kappa _{a}}\right) .
\end{equation}

For $m = 1$, the bounds (\ref{generalsumlimits}) become
\begin{equation}
\label{nsumlimits}
\frac{(1-\Delta _{a})\kappa _{b}}{(1+\Delta _{b})\kappa _{a}}<
n<\frac{(1+\Delta _{a})\kappa _{b}}{(1-\Delta _{b})\kappa _{a}}
\end{equation}
which using (\ref{Deltaapprox}) can be written as
\begin{equation}
\label{nsumlimitsapprox}
\frac{x}{8}<n<\frac{8x}{\epsilon ^{2}_{a}\epsilon _{b}^{2}}
\end{equation}
where
\begin{equation}
\label{xeqkaoverkb}
x=\frac{\epsilon ^{2}_{a}\kappa _{b}}{\kappa _{a}}.
\end{equation}

Let us first consider what happens to the \( n=1 \) mode as \( \kappa _{a} \)
approaches the limit in (\ref{kappaamax}). In this limit, \( \theta  \) becomes
\( \pi  \), while \( \omega  \) and \( \kappa _{a} \) approach fixed, finite
values. Thus \( k_{x}/\kappa _{a}\rightarrow 0 \), we can take the
small argument
limit of the Bessel function and the power in this mode approaches a finite
value given by (\ref{hindmarshpow}). When \( \kappa _{a} \) exceeds the limit
of (\ref{kappaamax}), the \( n=1 \) mode vanishes suddenly, causing a discontinuity
in the emitted power as a function of \( \kappa _{a} \). Similarly, if we let
\( \epsilon _{a} \) approach the limit given 
by \( \epsilon _{a}^{2}\approx 8\kappa _{a}/\kappa _{b} \),
we see that the power in the \( n=1 \) mode is again given by (\ref{hindmarshpow}),
which increases with \( \epsilon _{a}^{2} \) until it suddenly vanishes.

As \( \kappa _{a} \) decreases, the modes successively vanish (although only
the energy in the \( n=1 \) mode does so discontinuously), but the remaining modes that are
far from the limiting wave number grow larger. To see the overall effect, we
consider the case where \( \epsilon _{a}^{2}\kappa _{b}/\kappa _{a}\gg 1 \), or in terms 
of the amplitude $A_a \gg \sqrt{\lambda_a \lambda_b}$,
so that many different modes contribute to the total radiation.

In Appendix \ref{appendix:Besselapprox} we use the large-order
approximation for Bessel functions to compute the power
(\ref{powbess}) in this regime.  We find that the power declines
exponentially with $x$,
\begin{equation}
\label{coolestpowerever}
\frac{dP}{dz}\sim G\mu ^{2}\epsilon ^{2}_{b}\kappa _{a}e^{-\alpha x}
\end{equation}
where \( \alpha \approx 0.07867 \).  Thus there is essentially no
radiation from wiggles of significant amplitude and very different wavelengths.

\begin{figure}
{\centering \resizebox*{10cm}{7cm}{\includegraphics{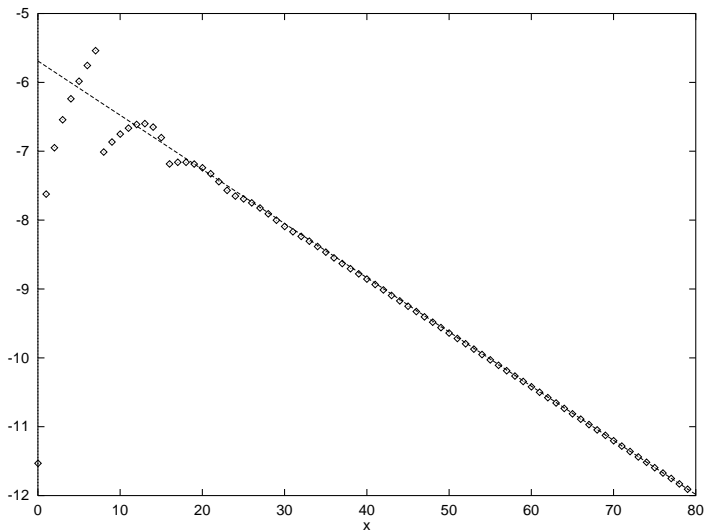}} \par}
\caption{Plot of \protect\( \ln \left( \frac{1}{G\mu ^{2}\kappa _{a}}
\frac{dP}{dz}\right) \protect \)
vs. \protect\( x=\epsilon ^{2}_{a}\kappa _{b}/\kappa _{a}\protect \) comparing
our analytic approximation (\ref{coolestpowerever},\ref{figurefit}) 
with \protect\( \beta =-5.69\protect \)
and \protect\( \alpha =-0.07867\protect \) (dashed line) 
with (\ref{powbess})
calculated numerically for \protect\( \epsilon _{a}=.1\protect \), 
\protect\( \epsilon _{b}=.05\protect \)
(points). }
\label{fig:fit}
\end{figure}

To compare the analytic approximation (\ref{coolestpowerever}) with our original
expression (\ref{powbess}), in Figure \ref{fig:fit} we show a plot of 
$\frac{dP}{dz}$
versus \( x=\epsilon ^{2}_{a}\kappa _{b}/\kappa _{a} \) with \( \epsilon _{a}=.1 \),
\( \epsilon _{b}=.05 \) and $\kappa_a$ fixed. The dashed line is 
\begin{equation}
\label{figurefit}
\ln \left( \frac{1}{G\mu ^{2}\kappa _{a}}\frac{dP}{dz}\right) =\alpha x+\beta
\end{equation}
with a best-fit value \( \beta =-5.69 \), and \( \alpha =-0.07867 \) as
above. The points are given by the sum over \( n \) and numerical
integration over \( \phi \) of (\ref{powbess}).  In the large $x$
regime, we see that the agreement with the analytic approximation is excellent.

The exponential suppression we have just found is valid when the 
Lorentz invariant conditions
\begin{equation}
\label{AbsAal}
A_{b}\ll \sqrt{\lambda _{a}\lambda _{b}}, A_{a}\gg \sqrt{\lambda _{a}\lambda _{b}}
\end{equation}
are satisfied. It is interesting to consider the effect of a Lorentz boost 
that makes both wiggle wavelengths similar. This boost makes
the condition (\ref{AbsAal}) for $A_a$ 
read $\epsilon_a \gg 1$. In this case 
we could not have performed the calculation at all because this is not the 
regime in which (\ref{Amusubnapprox}) is a valid approximation.
It must be true, however, because of Lorentz invariance, that the radiation 
is also exponentially suppressed.  
Although at first glance this may seem puzzling, it is
not too difficult to understand this result: In this limit, even though the wavelengths
are similar, the amplitude of one of the wave trains is much larger 
than the other ($A_a \gg A_b$) so that the larger wave-train appears 
to be traveling on an almost straight string and would not be 
expected to radiate \cite{Vachaspati}.

\section{Discussion and Conclusions}

We have calculated the power radiated in the form of gravitational waves from
oppositely moving excitation wave-trains in the small amplitude regime when
the wavelengths are comparable and when they are very different. We have shown
that small opposing excitations on strings will only radiate significantly if
their wavelengths are comparable. In this case we can always use (\ref{hindmarshpow})
for the power and the extra energy in the form of wiggles decreases as $1/t$
as discussed in the literature \cite{10}.

If, on the other hand, the wavelengths are very different then the
gravitational radiation can be suppressed.  If the amplitude $A_a$ of
the long-wavelength excitations is greater than the geometric mean
wavelength, $\sqrt{\lambda_a\lambda_b}$, then the radiation rate is
exponentially small.  Thus unless the amplitude of the long-wavelength
excitations is very small compared to their wavelength, there is
little interaction between modes of very different wavelengths.

The situation where the two wavelengths are very different occurs in
loops with chiral wiggles, i.e., when the wiggles are only on the
right- or left-moving excitations. In this case because the radiation
vanishes exponentially the wiggles may be very long-lived. We therefore
expect that if the loop is originally in a stable (non
self-intersecting) trajectory it will radiate and shrink until it
self-intersects, at the latest when the string loop is the same size
as the wiggles living on it. This may take a long time because the
tension on the strings is reduced by the presence of wiggles and the
motion may be slow \cite{11}.

When an excitation travels on an infinite string, it will eventually
be exposed to every possible sort of excitation going the other way,
and will thus eventually lose its energy by gravitational radiation.
However, when an excitation travels on a loop, it will encounter the same
oppositely moving excitations over and over again.  It is possible
that when a loop is formed there will be, by chance, an excess of
(say) right-going energy over left-going energy over a broad range of
wavelengths.  In this case, energy will be radiated until the
left-going excitations have been eliminated and some energy still
remains in the right-going ones.  Thus the string may become chiral
with respect to a certain range of wavelengths, and the chiral wiggles
in this range may then become long-lived.  This process is quite
complicated, and we will not try to analyze it here.  We merely note
that certain initial conditions may lead to long-lived chiral wiggles
over certain wavelength ranges, and we briefly investigate the
consequences of such excitations.

Almost all proposals for cosmic-string-induced astrophysical phenomena are related
to cusps. These are regions of the string that achieve extremely high Lorentz
boosts \cite{2}. Therefore here we will focus on the effects that the existence
of small wiggles may have on cusps.

Near a wiggly cusp the effective mass per unit length is
\begin{equation}
\label{mueff1}
\mu _{\rm eff}=\mu (1+\gamma ^{2}\epsilon ^{2})
\end{equation}
where \( \epsilon  \) is the amplitude to wavelength ratio of the wiggles
and $\gamma$ is the (transverse) Lorentz boost. If we think of the string as having 
a thickness due to the wiggles then before the cusp can be formed
the wiggles will typically overlap
when the gamma factor reaches the value \cite{KenandJose}
\begin{equation}
\label{gammaoverlap} 
\gamma_o \sim \sqrt{L/d}
\end{equation}
where $L$ is the size of the loop on which the cusp appears, 
$d \sim \epsilon / \kappa $ is the effective 
thickness due to the wiggles and $\kappa \sim 1/\lambda$ is the 
wiggle wavenumber. The cusp may also not form at all if 
there is sufficient back-reaction from the wiggles to deviate from
regular Nambu-Goto motion of the string, namely when the gamma factor in (\ref{mueff1}) 
reaches the value
\begin{equation} 
\label{gammabackreact}
\gamma_b \sim 1/\epsilon.
\end{equation}
Which of the two processes
is the dominant one depends on which one of the two gamma factors is 
the smallest, since it will be the one reached first by the cusp. 

It follows from this that there are two distinct wiggliness regimes. If $\gamma_b > \gamma_o$,
or \( L\kappa \epsilon < 1 \), there is not much change in the effective
mass of the string and the wiggles will not  significantly affect the
string motion. However, before the cusp can be formed, because
of the overlap of the wiggles, we will typically have a self-intersection that
chops off a loop of size \( l\sim \sqrt{\epsilon L/\kappa } \) which in turn
will fragment into about \( \sqrt{\kappa L/\epsilon } \) loops of the wiggle
size \( \epsilon /\kappa  \).

If, on the other hand, $\gamma_o > \gamma_b$, or \( L\kappa \epsilon > 1 \),
we are in the situation where the back-reaction is important. Much work has
gone into understanding the behaviour of strings in this regime \cite{12}. Typically 
one averages the string over scales larger than the wiggle size and
the effect is to increase the effective mass density and decrease the
tension of the string.
The situation is then remarkably similar to that of a superconducting string with a chiral
neutral current and
we expect the cusp in this case to be smoothed out and self-intersections to occur 
near it some of the time as discussed in \cite{13}. 

\begin{figure}
{\centering \resizebox*{10cm}{7cm}{\includegraphics{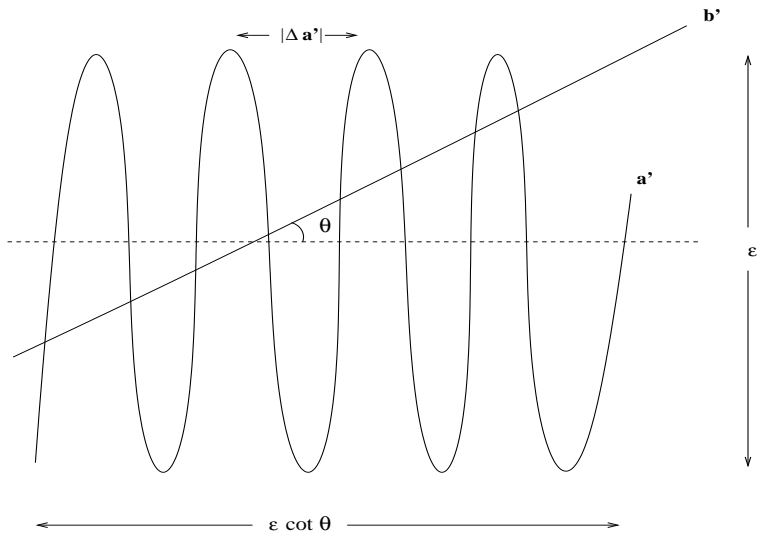}} \par}
\caption{Intersection on the unit sphere of a wiggly ${\bf a}'$ and a straight ${\bf b}'$. 
The dotted line resulting from averaging out the wiggles in ${\bf a}'$ makes an 
angle $\theta$ with ${\bf b}'$.}
\label{fig:wigglya}
\end{figure}

\begin{figure}
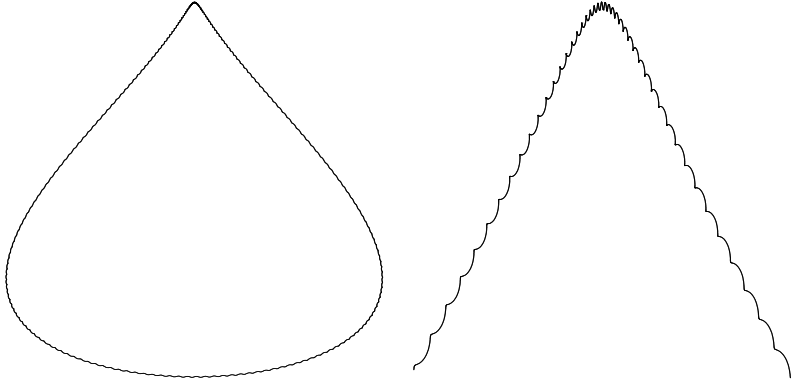

{\centering \begin{tabular}{cc}
\resizebox*{5cm}{5cm}{\includegraphics{101.eps}} &
\resizebox*{5cm}{5cm}{\includegraphics{201.eps}} \\
\end{tabular}\par}
\caption{A ``cusp'' in a wiggly string with $L\kappa\epsilon > 1$.  The wiggles have led the
overall shape of the string to be smoothed out.  Instead of one large
cusp, we expect
a large number of small cusps.}
\label{fig:cusp1}
\end{figure}

\begin{figure}
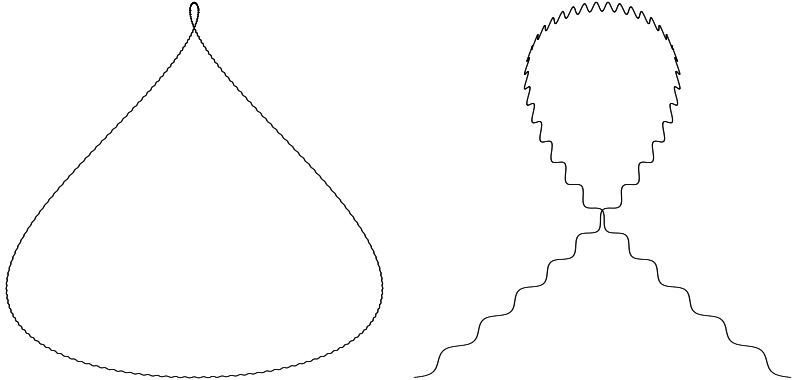

{\centering \begin{tabular}{cc}
\resizebox*{5cm}{5cm}{\includegraphics{301.eps}} &
\resizebox*{5cm}{5cm}{\includegraphics{401.eps}} \\
\end{tabular}\par}
\caption{Another possibility for a ``cusp''  in a wiggly string with
$L\kappa\epsilon > 1$.  In this case the smoothing effect of the
wiggles has led to a self-intersection.}
\label{fig:cusp2}
\end{figure}

We can try to estimate the number of cusps formed in both cases. If we consider 
the section of the unit sphere where a wiggly ${\bf a}'$ and a straight ${\bf b}'$ 
intersect, as in Figure \ref{fig:wigglya}, we can see that the number of crossings must 
be 
\begin{equation}
\label{nc1}
N_c \sim \epsilon \cot \theta / |\Delta {\bf a}'| 
\end{equation}
where $\epsilon \cot \theta $ is the length available for crossings in the section where
the ${\bf a}'$ and ${\bf b}'$ intersect on the unit sphere and $|\Delta {\bf a}'|$ is the distance
between peaks on the unit sphere. Typically
\begin{equation}
\label{deltaaprime}
|\Delta {\bf a}'| \sim |{\bf a}''| \Delta \sigma \sim \lambda/L
\end{equation}
where $\lambda$ is the wavelength of the wiggles and $L$ is the size of the loop. 
Taking $\cot \theta \sim 1$ gives
\begin{equation}
\label{nc2}
N_c \sim L \kappa \epsilon  
\end{equation}
for the number of cusps. It is interesting to note that this is also the quantity
that determines whether we are in the overlap or back-reaction regime. This means 
that in the overlap case
there will still be one cusp, just as in the case where there are no wiggles. 
As we have argued, however, 
self-intersections due to overlap
will take place before the cusp can be fully formed. 
In the back-reaction dominated case instead of a large cusp we will have 
about $L\kappa \epsilon$ small cusps on the smoothed out curve. 
In Figures \ref{fig:cusp1} 
and \ref{fig:cusp2} we plot pictures of Burden loops onto which 
we have superimposed chiral helical wiggles in the $L\kappa\epsilon >1$ case.

From this analysis it appears that wiggly cosmic string cusps may solve some
of the problems currently faced by cosmic-string-induced astrophysical phenomena.
Wiggly cusps could produce substantially more ultra-high energy cosmic
rays because of the presence
of many small cusps.  Gamma ray bursts may not be repeated because of
self-intersections near the cusp. Gravitational wave bursts may be more
frequent since, like cosmic rays, most of the energy comes from a small area
around the cusp.  Finally, wiggles may also lead to fine structure in
gravitational wave and gamma ray bursts.

Although some of these require different wiggliness regimes and would appear
mutually exclusive in fact they are not; we can ``have it both ways'' because
as we have shown modes of different wavelengths barely interact and as a 
consequence chirality can be independently established at different wavelengths.

\section{Acknowledgments}

We would like to thank Alex Vilenkin, J. J. Blanco-Pillado, Allen
Everett and Larry Ford for many useful discussions.  We would also like to
thank the referees of Nuclear Physics B for their careful reading of the manuscript
and useful suggestions. The work of KDO was partially funded by the NSF.

\appendix
\section{Periodic sources}\label{appendix:periodic}

In this appendix, we go from (\ref{weinberggravradforstrings}), which
gives the energy radiated from a localized interaction, to the power
per unit length for an infinite periodic source.  To do this, we
consider a wave train of length $L$ in $a$ and another of length $l$
in $b$, and then let $L$ and $l$ go to infinity.  The resulting energy will
be proportional to $Ll$, as shown below.  We will ignore any
contributions coming from the ends of the wave trains, because these
will not be proportional to $Ll$.

Thus we will take \( f \) to be an odd periodic function in \( u \), for
\( N_{a} \) periods of length \( \lambda _{a} \) (with $L =
N_a\lambda_a$) centered around \(
u=0 \), and $f = 0$ outside this range. Since \( f \) is odd, all
components of \( A^{\mu } \) will be real and
(\ref{weinberggravradforstrings}) becomes
\begin{equation}
\label{weinberggravradforstringssimple}
\frac{dE}{d\Omega }=\frac{G\mu ^{2}}{16\pi ^{2}}\int ^{\infty }_{0}d\omega 
\omega ^{2}\left| A\right| ^{2}\left| B\right| ^{2}.
\end{equation}

Because we are interested only in terms proportional to $Ll$, we
exclude the contribution of the straight parts of the string to the
integral in (\ref{amuofk}), and write this equation as
\begin{equation}
A^{\mu }=\sum _{j=-(N_{a}-1)/2}^{(N_{a}-1)/2}
\int _{(j-1/2)\lambda _{a}}^{(j+1/2)\lambda _{a}}du\, 
a'^{\mu }e^{ik\cdot a(u)/2}\,.
\end{equation}
Now $a'$ is periodic, and so $a (u+j\lambda_a)-a (u)$ is a constant
independent of $u$.  Thus
\begin{equation}
\label{Amuofkplus}
A^\mu=\sum _{j=-(N_{a}-1)/2}^{(N_{a}-1)/2}e^{ijK_{+}\lambda _{a}/2}
\int _{-\lambda _{a}/2}^{\lambda _{a}/2}du\, a'^{\mu }e^{ik\cdot a(u)/2}
\end{equation}
where
\begin{eqnarray}
\label{Kplus}
K_{+}&=&\frac{k\cdot (a(\lambda _{a})-a(0))}{\lambda _{a}}
=
\frac{1}{\lambda _{a}}\int _{0}^{\lambda _{a}}du\, 
k\cdot a'(u)
\nonumber
\\
&=&\omega +k_{z}\Delta _{a}
=\omega (1+\Delta _{a}\cos \theta )
\end{eqnarray}
with 
\begin{equation}
\Delta _{a}=\frac{1}{\lambda _{a}}\int _{0}^{\lambda _{a}}du\, \sqrt{1-f'^{2}.}
\end{equation}

The sum in (\ref{Amuofkplus}) is 
\begin{equation}
\label{sumeqsins}
\sum _{j=-(N_{a}-1)/2}^{(N_{a}-1)/2}e^{ijx}=\frac{\sin (N_{a}x_{a}/2)}{\sin (x_{a}/2)}
\end{equation}
with \( x_{a}=\lambda _{a}K_{+}/2 \). This function has peaks of height \( N_{a} \)
whenever \( x_{a} \) is an integer multiple of \( 2\pi  \) and in the vicinity
of these peaks the function is effectively \( N_{a}\sinc(N_{a}x_{a}/2\pi ) \).
However, because what appears in (\ref{weinberggravradforstringssimple}) is
the square of \( A^{\mu }(k) \) what we actually need to evaluate 
is \( N^{2}_{a}\sinc^{2}(N_{a}x_{a}/2\pi ) \).
In the large \( N_{a} \) limit \( N_{a}\sinc^{2}(N_{a}y)\! \rightarrow \! \delta (y) \)
so we can write its square as 
\begin{equation}
\label{sinseqdeltasums}
\frac{\sin ^{2}(N_{a}x/2)}{\sin ^{2}(x/2)}\approx 
N_{a}\sum ^{\infty }_{n=-\infty }\delta (\frac{\lambda _{a}K_{+}}{2(2\pi )}-n)
=2\kappa _{a}N_{a}\sum _{n=-\infty }^{\infty }\delta (K_{+}-2n\kappa _{a}).
\end{equation}

A similar set of expressions can be written for \( B^{\mu }(k) \) which allow
(\ref{weinberggravradforstringssimple}) to be re-written as 
\begin{equation}
\label{weinberggravradforstringsmoresimple}
\frac{dE}{d\omega d\Omega }=
\frac{G\mu ^{2}}{4\pi ^{2}}\kappa _{a}\kappa _{b}N_{a}N_{b}\sum _{n,m}
\omega ^{2}\delta (K_{+}-2n\kappa _{a})\delta (K_{-}-2m\kappa _{b})
{\cal A} ^{2}{\cal B} ^{2}
\end{equation}
where
\begin{equation}
\label{alphabeta}
{\cal A} =\int _{-\lambda _{a}/2}^{\lambda _{a}/2}du\, a'^{\mu }e^{ik\cdot a(u)/2},
\: {\cal B} =\int _{-\lambda _{b}/2}^{\lambda _{b}/2}dv\, b'^{\mu }e^{ik\cdot b(v)/2}.
\end{equation}

All the variables with the \( b \) subscript have the obvious meaning and 
\( K_{-}=\omega (1-\Delta _{b}\cos \theta ) \).
The sum in (\ref{weinberggravradforstringsmoresimple}) is performed for positive
values of \( n \) and \( m \) only since \( K_{+} \) and \( K_{-} \) are
nonnegative.

We can now use the \( \delta\) functions to perform the integral 
in (\ref{weinberggravradforstringssimple}) over \( \omega  \) and to integrate over \( \theta  \). We do 
this by a suitable change of variables,
namely, 
\begin{equation}
\label{omandcosthetaofKandDelta}
\omega =(\Delta _{b}K_{+}+\Delta _{a}K_{-})/(\Delta _{a}+\Delta _{b}),\: \cos \theta 
=(K_{+}-K_{-})/(\Delta _{b}K_{+}+\Delta _{a}K_{-}))
\end{equation}
so that
\begin{equation}
\label{domdcostheta}
d\omega d\cos \theta =
dK_{+}dK_{-}/(\Delta _{b}K_{+}+\Delta _{a}K_{-})=
dK_{+}dK_{-}/(\omega (\Delta _{a}+\Delta _{b})).
\end{equation}

Upon integration we get
\begin{equation}
\label{dEdphi}
\frac{dE}{d\phi }
=\frac{G\mu ^{2}}{\Delta _{a}+\Delta _{b}}Ll\sum _{n,m}\omega |A_{n}|^{2}|B_{m}|^{2}
\end{equation}
where the vector
\begin{equation}
A^{\mu }_{n}=\frac{1}{\lambda _{a}}\int _{-\lambda _{a}/2}^{\lambda _{a}/2}du\, 
a'^{\mu }(u)e^{ik\cdot a(u)/2}\,.
\end{equation}
and there is a similar expression for \( B_{n}^{\mu } \). Dividing (\ref{dEdphi})
by the volume of the world sheet where the interaction occurs, \( Ll/2 \), yields the power per unit length per unit
azimuthal angle \( \phi  \) 
\begin{equation}
\label{dPdzdphi}
\frac{dP}{dzd\phi }=\frac{G\mu ^{2}}{\Delta _{a}+\Delta _{b}}
\sum _{n,m}(n\kappa _{a}+m\kappa _{b})|A_{n}|^{2}|B_{m}|^{2}
\end{equation}

Keeping the above in mind we can see that the significance of the appearance
of the \( \delta \) functions is that for periodic sources the radiation is
only emitted in discretized directions and frequencies. Explicitly, for a set
of given \( \Delta _{a} \), \( \Delta _{b} \), \( \kappa _{a} \) and \( \kappa _{b} \)
we have that
\begin{eqnarray}
\omega &=&2\frac{n\kappa _{a}\Delta _{b}+m\kappa _{b}\Delta _{a}}{\Delta _{a}
+\Delta _{b}}\,,\\
\cos \theta &=&
\frac{n\kappa _{a}-m\kappa _{b}}{n\kappa _{a}\Delta _{b}+m\kappa _{b}\Delta _{a}}\,.\label{costhetaappendix}
\end{eqnarray}
This discreteness in frequencies and angles
arises from the temporal and spatial periodicity of the system.

If we had \( \Delta _{a}=\Delta _{b}=1 \),
then the right hand side of (\ref{costhetaappendix}) would always lie between
-1 and 1. However, for waves of finite amplitude, this is not the case and thus
not all modes are present. This limitation restricts the sum over \( n \) and
\( m \) to satisfy
\begin{equation}
(1-\Delta _{a})/(1+\Delta _{b})<|n\kappa _{a}/m\kappa _{b}|<(1+\Delta _{a})/(1-\Delta _{b}).
\end{equation}

\section{Large order approximation}
\label{appendix:Besselapprox}

In this appendix, we compute an analytic approximation to the
radiation power (\ref{powbess}) using the Bessel function
approximation for large order \cite{9}
\begin{equation}
\label{bessapprox}
J_{n}(n\sech\alpha )\approx \frac{e^{n(\tanh \alpha -\alpha )}}
{\sqrt{2\pi n\tanh \alpha }}
\end{equation}
which is valid when $\tanh\alpha > 1/n$.

Using (\ref{kx}) we can see that
\begin{eqnarray}
\label{kxoverkasq}
\frac{k^{2}_{x}}{\kappa ^{2}_{a}}&=&\frac{\cos ^{2}(\phi )}{4}\left\{ (\Delta _{b}n
+\Delta _{a}x/\epsilon ^{2}_{a})^{2}-(n-x/\epsilon ^{2}_{a})^{2}\right\} 
\nonumber
\\
&\approx& \frac{\cos ^{2}(\phi )}{4}\left\{ 4nx/\epsilon ^{2}_{a}
-\frac{x^{2}}{2\epsilon ^{2}_{a}}-\frac{\epsilon ^{2}_{b}}{2}n^{2}\right\} 
\end{eqnarray}
which allows us to express the square of the argument of the remaining Bessel
function in (\ref{powbess}) as
\begin{equation}
\label{besselarg}
\frac{k^{2}_{x}\epsilon _{a}^{2}}{4\kappa ^{2}_{a}}
\approx \frac{\cos ^{2}(\phi )}{16}\left\{ 4nx-\frac{x^{2}}{2}
-\frac{\epsilon ^{2}_{a}\epsilon ^{2}_{b}}{2}n^{2}\right\} \equiv
y^{2}
\end{equation}
and therefore
\begin{equation}
\label{powbessy}
\frac{dP}{dzd\phi }\approx \frac{G\mu ^{2}\epsilon ^{2}_{a}\epsilon _{b}^{2}}{8}
\sum _{n}(n\kappa _{a}+\kappa _{b})\left( \frac{n^{2}}{y^{2}}
-\frac{1}{2}\right) J^{2}_{n}\left( y\right) .
\end{equation}

Bearing in mind (\ref{bessapprox}) we  take 
\begin{equation}
\label{zeqn}
\sech\alpha =w=y/n\approx \cos (\phi )\{4x/n-x^{2}/2n^{2}\}^{1/2}/4
\end{equation}
so that we can write
\begin{equation}
\label{powbessz}
\frac{dP}{dzd\phi }\approx \frac{G\mu ^{2}\epsilon ^{2}_{a}\epsilon _{b}^{2}}{8}
\sum _{n}(n\kappa _{a}+\kappa _{b})\left( \frac{1}{w^{2}}-\frac{1}{2}\right) 
J^{2}_{n}\left( nw\right) .
\end{equation}

In our case it is easy to see that \( w \) has a maximum at \(
\tilde{n}_{m}=x/4 \) and the value of the function at this peak is \(
\tilde{w}_{m}=\cos (\phi )/\sqrt{2} \).  This means that $w$ is at most \( 1/\sqrt{2} \) and thus $\tanh\alpha$ is at least
$1/\sqrt{2}$, and we see that (\ref{bessapprox}) is indeed a good
approximation.

We will now perform the sum over \( n \). To do the calculation analytically we must
first approximate the sum by an integral,
which we can do provided the values of the function are sufficiently close to
one another in the region in which we are performing the sum. Secondly, we will
assume that the largest contribution to the Bessel function comes from a maximum
and that we can approximate that integral by multiplying the height at the peak
times the width of this peak at the maximum 
\begin{equation}
\label{superapprox}
\frac{dP}{dzd\phi }\sim \sum _{n}J_{n}^{2}(nw)\sim \int dnJ_{n}^{2}(nw)
\sim J^{2}_{max}\Delta n.
\end{equation}
This will be a reasonable approximation provided the function is sufficiently
smooth around the maximum. Below we show that this condition is satisfied.

We can take
\begin{equation}
\label{zsq}
w^{2}\approx c^{2}\frac{x}{2n}\left( 1-\frac{x}{8n}\right) 
\end{equation}
where 
\begin{equation}
\label{ccosphi}
c=\cos \phi /\sqrt{2}
\end{equation}
If we let 
\begin{equation}
\label{pxn}
p=\frac{x}{4n}-1
\end{equation}
then 
\begin{equation}
\label{zsqpsqcsq}
w^{2}=c^{2}(1-p^{2})
\end{equation}
which obviously has the maximum value \( c^{2} \) at \( p=0 \). To find the
maximum of the integrand of (\ref{superapprox}) we use the very rough approximation
\begin{equation}
\label{lnJ}
\ln J_{n}(nw)\sim n(\tanh \alpha -\alpha )=
n(\sqrt{1-w^{2}}-\sech^{-1}w)\,.
\end{equation}

Now if the factor of \( n \) were not present, we would reason that since
\( w=\sech\alpha  \)
is a decreasing function of \( \alpha  \), and \( \tanh \alpha -\alpha  \)
is also a decreasing function of \( \alpha  \), then the right hand side not
counting the \( n \) is an increasing function of \( w \), and (\ref{lnJ}) will
have its maximum where \( w \) has its maximum. Since the right hand side is
negative, and also a decreasing function of \( n \), we can infer that
the actual value of \( n \) that maximises (\ref{lnJ}) is actually less than the one that 
maximises \( w \). We can thus write 
\begin{equation}
\label{nxpc}
n=\frac{x}{4(1+p)}=\frac{x}{4(1+\sqrt{1-w^{2}/c^{2}})}
\end{equation}
and 
\begin{equation}
\label{lnjsimsech}
\ln J_{n}(nw)\sim 
\frac{x}{4}\frac{\sqrt{1-w^{2}}-\sech^{-1}w}{1+\sqrt{1-w^{2}/c^{2}}}\,.
\end{equation}
Setting the derivative of (\ref{lnjsimsech}) to zero gives
\begin{equation}
\label{nastycondition}
\frac{1-w^{2}-\sqrt{1-w^{2}}\sech^{-1}w}{w^{-1}-w}
=-\frac{1-w^{2}/c^{2}+\sqrt{1-w^{2}/c^{2}}}{w/c^{2}}.
\end{equation}

\begin{figure}
{\centering \resizebox*{0.6\columnwidth}{0.35\textheight}{\includegraphics{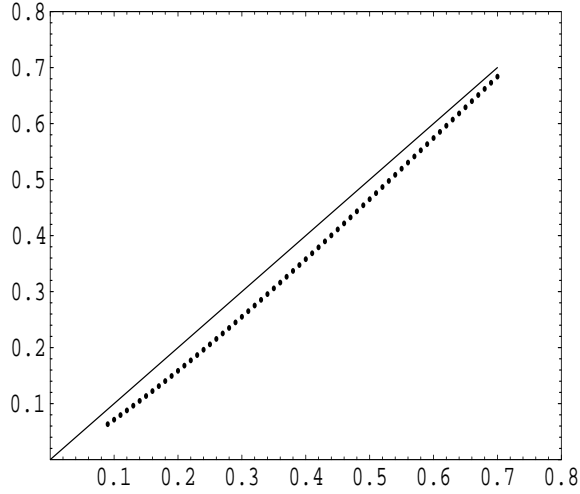}} \par}
\caption{
Plot of $w_m$ (points) on the $y-$axis, as a 
function of $c=\cos ( \phi )/\sqrt{2}$. The solid line is $\tilde{w}_m=\cos(\phi)/\sqrt{2}$ 
for comparison.}
\label{fig:zm}
\end{figure}

This equation can be solved (although not analytically) for \( w_{m} \), the
value of \( w \) that maximises (\ref{lnJ}) for a particular \( c \) and from it, 
using (\ref{nxpc}),
one can obtain \( n_{m} \), the value of \( n \) at the maximum of (\ref{lnjsimsech}).
Note that both values will be in general different from \( \tilde{n}_{m} \)
and \( \tilde{w}_{m} \), the maximum of \( n \) and \( w \) at the maximum
of the function \( w \). Figure \ref{fig:zm} shows a plot of $w_m$, the maximum of (\ref{lnJ}), 
as a function of $c=\cos ( \phi )/\sqrt{2}$. As expected the value of $w$ where
(\ref{lnJ}) is at a maximum is slightly less than the maximum value of $w$.

In order to estimate the width at the peak we can again use (\ref{lnJ}). However
in this case we can use the width at the maximum of the function \( w
\), i.e., at
\( p=0 \), or \( \tilde{n}_{m}=x/4 \), because it is sufficiently close to
the maximum of \( J_{n}(nw) \). This simplifies the
calculation considerably. From (\ref{lnJ}) one can see that
\begin{equation}
\label{lnjsimntamhalpha}
\frac{d}{d\alpha }\ln J\sim -n\tanh ^{2}\alpha 
\end{equation}
and we can compute the width using the Taylor expansion
\begin{equation}
\label{lnjlnj0}
\ln J\sim \ln J_{0}-\Delta \alpha n\tanh ^{2}\alpha .
\end{equation}
Taking \( \ln J-\ln J_{0}\sim -1 \) we find 
\begin{equation}
\label{deltaalpha}
\Delta \alpha \sim -\frac{1}{n\tanh ^{2}\alpha }.
\end{equation}

We need to express our answer in terms of \( \Delta n \) however so we take
\begin{equation}
\label{deltazsechtanh}
\Delta w\sim \frac{\partial w}{\partial \alpha }\Delta \alpha =
\sech\alpha \tanh \alpha \Delta \alpha =\frac{\tilde{w}_{m}}
{\tilde{n}_{m}\sqrt{1-\tilde{w}_{m}^{2}}}
\end{equation}
and 
\begin{equation}
\label{deltaz}
\Delta w=\frac{\partial ^{2}w}{\partial n^{2}}\frac{(\Delta n)^{2}}{2}\,,
\end{equation}
because the first derivative is \( 0 \) at the maximum, yielding
\begin{equation}
\label{deltan}
\Delta n=\left( \frac{\tilde{n}_{m}}{\sqrt{1-\tilde{w}_{m}^{2}}}\right) 
^{1/2}\sim \tilde{n}^{1/2}_{m}.
\end{equation}

As advertised above, 
since \( \tilde{n}^{1/2}_{m}\propto (\epsilon ^{2}_{a}\kappa _{b}/\kappa _{a})^{1/2} \)
is large in the small \( \kappa _{a} \) limit, the function around the maximum
is sufficiently smooth that we can now perform the sum by approximating it as
an integral and further approximating this integral by multiplying the value
of the function at the peak by the width near the peak \( \Delta n \). 

For the power per unit length per unit azimuthal angle this procedure yields
\begin{equation}
\label{powbesszapprox1}
\frac{dP}{dzd\phi }\sim \frac{G\mu ^{2}\epsilon ^{2}_{a}\epsilon _{b}^{2}}{8}
(n_{m}\kappa _{a}+\kappa _{b})\left( \frac{1}{w_{m}^{2}}-\frac{1}{2}\right) 
J^{2}_{n_{m}}\left( n_{m}w_{m}\right) \Delta n\,,
\end{equation}
where \( n_{m} \) and \( w_{m} \) are given above. 

The integral over \( \phi  \) can be approximated in an analogous way,
taking the value of the Bessel function at the peak, which is at \( \phi =0 \),
and finding the width at this peak. Evaluating the function at the maximum is
straightforward, one simply needs to solve (\ref{nastycondition}) for \( \cos \phi =1 \),
or equivalently \( c=1/\sqrt{2} \). This procedure yields \( w_{0}\approx .691584 \)
and from (\ref{nxpc}), \( n_{0}\approx x/4.83352 \).

In this case the width can be found by using \( \ln J\sim f(w_{m}) \) as given
by (\ref{lnjsimsech}) and we can see that
\begin{equation}
\label{d2fdphisq}
\frac{d^{2}f(w_{m})}{d\phi ^{2}}
=\frac{df}{dw_{m}}\frac{d^{2}w_{m}}{d\phi ^{2}}
\end{equation}
because \( dw_{m}/d\phi =0 \).
Using \( \ln J-\ln J_{0}\sim -1 \) again yields
\begin{equation}
\label{deltaphisqdfdzm}
\frac{(\Delta \phi )^{2}}{2}\frac{df}{dw_{m}}\frac{d^{2}w_{m}}{d\phi ^{2}}\sim -1
\end{equation}
and because \( f\propto x \), ignoring factors of \( O(1) \) we have that
\begin{equation}
\label{deltaphi}
\Delta \phi \sim x^{-1/2}.
\end{equation}

These considerations yield for the power per unit length
\begin{eqnarray}
\label{powbesszapprox2}
\frac{dP}{dz}&\sim& G\mu ^{2}\epsilon ^{2}_{a}\epsilon ^{2}_{b}(n_{0}\kappa _{a}
+\kappa _{b})J^{2}_{n_{0}}\left( n_{0}w_{0}\right) \Delta n\Delta \phi 
\nonumber
\\
&\sim  &
G\mu ^{2}\epsilon ^{2}_{a}\epsilon ^{2}_{b}(n_{0}\kappa _{a}+\kappa _{b})
J^{2}_{n_{0}}\left( n_{0}w_{0}\right) 
\end{eqnarray}

Expressing this equation instead using (\ref{bessapprox}), with
inverse hyperbolic functions written out explicitly,
and ignoring factors of \( O(1) \) yields
\begin{equation}
\label{powzmax}
\frac{dP}{dz}\sim G\mu ^{2}\epsilon ^{2}_{b}\kappa _{a}
\left( \frac{w_{0}e^{\sqrt{1-w^{2}_{0}}}}{1+\sqrt{1-w^{2}_{0}}}\right) ^{2n_{0}}
\end{equation}
which can be written more intelligibly as
\begin{equation}
\frac{dP}{dz}\sim G\mu ^{2}\epsilon ^{2}_{b}\kappa _{a}e^{-\alpha x}
\end{equation}
where \( \alpha \approx 0.07867 \) can be calculated from \( w_{0} \) and
\( n_{0} \).

\end{document}